\documentclass[preprint,showpacs,preprintnumbers,amsmath,amssymb]{revtex4}

\usepackage{graphicx}
\usepackage{dcolumn}
\usepackage{bm}

\begin{document}

\preprint{APS/123-QED}

\title{First-order superfluid-Mott insulator transition of spinor bosons 
in an optical lattice}

\author{Takashi Kimura}
 \affiliation{Department of Information Sciences, Kanagawa University, 2946 Tsuchiya, Hiratsuka, Kanagawa 259-1293, Japan}
 \email{tkimura@info.kanagawa-u.ac.jp}
\author{Shunji Tsuchiya and Susumu Kurihara}
\affiliation{Department of Physics, Waseda University, 3-4-1 Ohkubo, Shinjuku, Tokyo 169-8555, Japan
}%

\date{\today}

\begin{abstract}
We study the superfluid-Mott insulator transition 
of antiferromagnetic spin-1 bosons in an optical 
lattice described by a Bose-Hubbard model. 
Our variational study with the Gutzwiller-type trial wave function 
determines that the superfluid-Mott insulator transition is a first-order one 
at a part of the phase boundary curve, contrary to the spinless case. 
This first-order transition may be observed through an experiment, such 
as a Stern-Gerlach type, under a magnetic field. 
\end{abstract}

\pacs{03.75.Lm, 03.75.Mn, 03.75.Hh, 32.80.Pj}
\maketitle
Superfluid (SF) transition is one of the most 
striking phenomena of condensed matter physics.
In particular, critical phenomena of superfluid transition, 
including the order of the transition,  
have been extensively studied for several decades.
The quantum superfluid-Mott insulator (SF-MI) transition has been 
studied in granular superconductors \cite{Goldman}, Josephson-junction
arrays \cite{Fazio}, and helium absorbed in the porous media \cite{Reppy}.
Recently, the SF-MI transition of bosons in an optical lattice
has been very clearly observed \cite{Greiner}. 
Jaksch {\it et al.} \cite{Jaksch} have shown that bosons 
in an optical lattice can be described 
by a Hubbard model \cite{M.P.A.Fisher} (a Bose-Hubbard model).   
The Bose-Hubbard model for spinless bosons 
has been theoretically studied 
for the last two decades \cite{M.P.A.Fisher,QMC,BoseGutz,Oosten,Sheshadri}.
Monte Carlo studies \cite{QMC} have confirmed that the 
transitions of the clean and dirty Bose-Hubbard models of spinless bosons
are continuous as suggested by analytical studies \cite{M.P.A.Fisher}.

It is also interesting to study the Bose-Hubbard model of spinor bosons \cite{spinor}. 
Demler and Zhou \cite{DemlerZhou} have discussed 
several unique properties of spin-1 bosons in an optical lattice. 
In a previous paper \cite{Tsuchiya},
we determined the SF-MI phase boundary 
of spin-1 bosons with an antiferromagnetic interaction
using a perturbative mean-field approximation (PMFA) \cite{Oosten}, 
which gives a phase boundary close to that obtained by Monte Carlo 
studies for the case of spinless bosons.

An excellent trial wave function for studying the Bose-Hubbard model 
is a Gutzwiller-type wave function (GW) \cite{Gutzwiller},  
which has been frequently used for the Hubbard model for electrons \cite{Yokoyama}.
For spinless bosons, the GW describes a second-order 
SF-MI transition and obtains a phase boundary curve, 
which is in an exact agreement with that obtained using the PMFA \cite{Oosten}.
A GW for spinor bosons has been employed only recently
for a non-uniform system \cite{Yamashita}.

In the present study, by employing the GW, we show the SF-MI transition 
can be a first-order one at a part of the phase boundary. 
The first-order SF-non-SF transition is rare and interesting. 
For example, as stated above, the SF-MI transition of the spinless bosons 
is second-order one \cite{M.P.A.Fisher}. 
Hence, the spin degree of freedom has an essential 
role in the first-order transition. 
The first-order transition can be observed by experiments,  
such as Stern-Gerlach type, under a magnetic field. 

The Bose-Hubbard Hamiltonian \cite{DemlerZhou,nematic1,note1} 
of spin-1 bosons is given by $H=H_0+H_1$,
\begin{eqnarray}
H_0&=&-t\sum_{\langle i,j \rangle}\sum_\alpha(a_{i \alpha}^\dagger a_{j \alpha}^{}
 + a_{j \alpha}^\dagger a_{i \alpha}^{}), \nonumber\\
H_1&=&-\mu\sum_i\sum_\alpha a_{i \alpha}^\dagger a_{i \alpha}^{} 
+\frac{1}{2}U_0\sum_i\sum_{\alpha,\beta} a_{i \alpha}^\dagger a_{i \beta}^\dagger
a_{i \beta}^{}a_{i \alpha}^{}\nonumber\\
&&+\frac{1}{2} U_2 \sum_i\sum_{\alpha,\beta,\gamma,\delta} 
a_{i \alpha}^\dagger a_{i \gamma}^\dagger
{\bf F}_{\alpha \beta} \cdot {\bf F}_{\gamma \delta} a_{i \delta}^{}
 a_{i \beta}^{}.\label{H}\nonumber\\
&=&\sum_i[-\mu\hat{n}_i+\frac{1}{2}U_0\hat{n}_i(\hat{n}_i-1)
+\frac{1}{2}U_2({\hat{\mathbf S}_i}^2-2\hat{n}_i)].
\end{eqnarray}
Here, $\mu$ and $t$ are the chemical potential and the
hopping matrix element between adjacent sites, respectively.  
$U_0$ and $U_2$ represent 
the spin-independent and the spin-dependent interactions between bosons, respectively. 
We assume an antiferromagnetic interaction ($U_2>0$).  
$a_{i \alpha}$ and $a_{i \alpha}^\dagger$ 
are the annihilation and creation 
operators, respectively, for a boson at site $i$ with spin magnetic quantum number 
$\alpha=1,0,-1$. $n_i\equiv\sum_\alpha n_{i\alpha}$ ($n_{i\alpha}\equiv 
a_{i \alpha}^\dagger a_{i \alpha}$) is a number operator at site $i$. 
$ {\hat{\bf S}_i}\equiv \sum_{\alpha,\beta}a_{i \alpha}^\dagger 
{\bf F}_{\alpha \beta} a_{i \beta}$ is a spin operator at site $i$ and  
${\bf F}_{\alpha \beta}$ represent the spin-1 matrices. 
$\langle i,j \rangle$ expresses a summation 
for all the sets of adjacent sites.

The GW of the model is defined 
as $\Psi\equiv \prod_i\Phi_i$. Here,
$\Phi_i$ is a wave function at site $i$ but  
the functional form of $\Phi_i$ is assumed to be 
site-independent such that $\Phi_i=\Phi$. 
$\Phi$ is written as a linear combination of states with 
$N$ bosons at a site as 
$\Phi=\sum_{N}g(N)|N\rangle$, 
where $|2n+1\rangle=\sum^{2n+1}_{S=1}f(2n+1,S)|2n+1,S\rangle$ and 
$|2n\rangle=\sum^{2n}_{S=0}f(2n,S)|2n,S\rangle$; 
$|N,S\rangle$ is the state where 
$N$ is the number of bosons and $S$ is 
the total spin, where 
$S$ must be odd for an odd and even for an even $N$ \cite{DemlerZhou}.
We assume that ${\hat S}_z|N,S\rangle=0$ \cite{note2}. 
Hence, $\Phi$ is an eigenstate of $S_z$ (not $S$) as 
a quantum spin nematic state \cite{nematic1} in the MI state.
$\Phi$ can interpolate between two limits about $U_2$ as 
$\Phi=\sum_n[g(2n)|2n,S=0\rangle+g(2n+1)|2n+1,S=1\rangle]$ 
($U_2\rightarrow\infty$) that minimizes the antiferromagnetic interaction 
$H_{\rm AF}\equiv\frac{1}{2}\sum_i
U_2({\hat{\mathbf S}_i}^2-2\hat{n}_i)$ 
and $\Phi=\sum_n g(N) a_0^{\dagger N}|0\rangle/\sqrt{N!}$
($U_2\rightarrow 0$) that includes high-spin states and 
minimizes the kinetic energy, 
where $|0\rangle$ is the vacuum of bosons. 
We note that the latter GW for $U_2=0$ has the same form as 
the GW of the spinless bosons \cite{BoseGutz}.
We numerically optimize the variational 
parameters $g(N)$ and $f(N,S)$ to minimize the 
energy expectation value by Powell's method \cite{numerical}
under the normalization conditions $\sum_{N} |g(N)|^2=1$ 
and $\sum_S |f(N,S)|^2=1$. 
We select the states where the number of bosons range from $N=0$ to $N=6$, 
which are sufficient for a numerical convergence in 
the parameter regime studied in this paper.
We define that the MI phase has a 
zero particle number fluctuation, and the SF 
phase has a finite particle number fluctuation \cite{note0}.
In an SF phase close to an MI one with $N$ bosons, 
probability densities of the states for a different values of $N$
can be considered as SF order parameters. 

\begin{figure}
\includegraphics[height=2.5in]{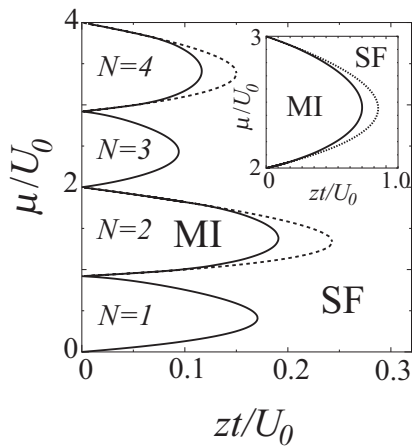}
\caption{
Phase diagram of the Bose-Hubbard model of spin-1 bosons for
$U_2/U_0=0.04$. 
Here, $z$ is the number of adjacent sites in the lattice. 
SF and MI indicate the superfluid and the Mott-insulating phases, 
respectively. The solid and dashed curves 
are obtained using the GW and the PMFA, respectively. 
The inset indicates the SF-MI phase boundary around the 
MI state with $N=3$ for $U_2/U_0=0.001$. 
}
\label{phase}
\end{figure}

Figure \ref{phase} shows the phase diagram for $U_2/U_0=0.04$,  
which corresponds to ${}^{23}{\mathrm Na}$ 
atoms \cite{spinor}.
The solid and dashed curves indicate the SF-MI phase boundaries 
using the GW and the PMFA, respectively. 
Here, $z$ is the number of adjacent sites in the lattice. 
Interestingly, at a part of the phase boundary curves, 
the GW slightly redefines the phase boundary curves obtained using the PMFA.  
It will be important to note that for spinless bosons, 
the phase boundary obtained using the GW is the same 
as that obtained using the PMFA. 
However, an even-odd conjecture predicted in Ref. \cite{Tsuchiya}
still clearly holds; the MI phase with an even $N$ is 
strongly stabilized against the SF phase.

On the other hand, in Fig. \ref{phase}, the SF-MI phase boundary 
around the MI phase with an odd $N$ obtained using
the GW is the same as that obtained using the PMFA.
This agreement always holds around the MI phase with $N=1$.
However, if we assume a much smaller $U_2$, we see a similar 
discrepancy between the two methods (inset of Fig. \ref{phase})
around the MI phase with $N=3$. 

It should be noted that the GW including only a set of low-spin states 
exactly reproduces the phase boundary obtained using the PMFA. 
For an even $N=2n$, assuming $g(2n\pm 1)=\epsilon_{2n\pm1}$, 
$g(2n)=\sqrt{1-\epsilon_{2n+1}^2-\epsilon_{2n-1}^2}$, and 
$f(2n\pm1,S=1)=f(2n,S=0)=1$ ($\epsilon_{2n\pm1}$ are infinitesimal), 
we analytically reproduce the phase boundary curve around the 
Mott phase with $N=2n$ obtained using the PMFA (Eq. 30 in Ref. \cite{Tsuchiya}). 
We also reproduce the phase boundary obtained using the PMFA  
around the Mott phase with an odd $N=2n+1$ by numerical optimization
of the GW only including the states $|2n+1,S=1\rangle$, 
$|2n+2,S=0\rangle$, $|2n+2,S=2\rangle$, $|2n,S=0\rangle$, 
and $|2n,S=2\rangle$.
These sets of the low-spin states are 
nothing but the states that emerge as zero-order states 
or intermediate states in the second-order 
PMFA which determines the phase boundary \cite{Tsuchiya}. 
This is consistent with the case of the phase boundary
around the Mott state with $N=1$ and that of spinless bosons.  

\begin{figure}
\centerline{\includegraphics[height=2.4in]{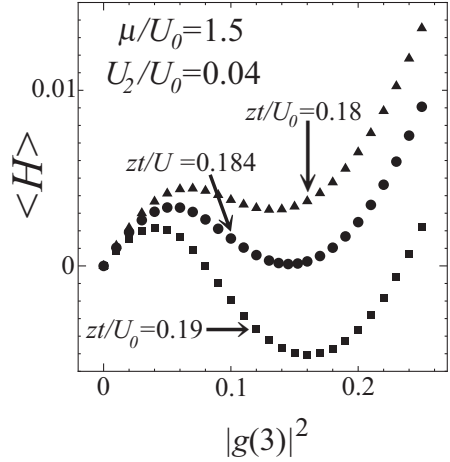}}
\caption{Expectation values of the total energy 
per site $\langle H\rangle$ as a function of 
$|g(3)|^2$ for $U_2/U_0=0.04$ and $\mu/U_0=1.5$. 
The other variational parameters are determined to 
minimize the energy. The origin of the vertical axis 
corresponds to the energy of the MI state with $N=2$.
$zt=1.9$,  $zt=1.84$, and $zt=1.8$ correspond 
to an MI state, a state very close to the phase boundary, 
and an SF state, respectively.  
}
\label{first}
\end{figure}

\begin{figure}
\centerline{\includegraphics[height=2.8in]{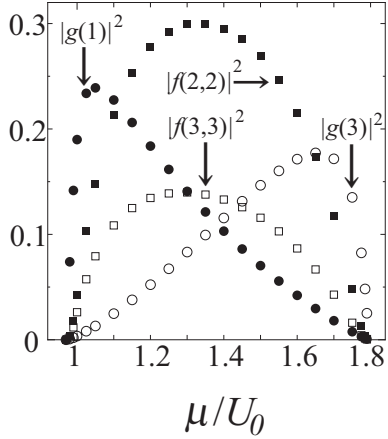}}
\caption{Variational parameters just on the phase boundary between 
the SF and  MI phases with $N=2$  for $U_2/U_0=0.04$.
The black and white circles indicate $|g(1)|^2$ and $|g(3)|^2$, respectively  
and the black and white squares 
indicate $|f(2,2)|^2$ and $|f(3,3)|^2$, respectively. 
}
\label{boundary}
\end{figure}

In our GW, the SF phase has a polar symmetry ($\langle {\bf S}\rangle=0$) 
and not only does the lowest spin state ($S=1$ or $S=0$)
at a given $N$ but also higher spin states have finite probability 
densities. 
The probability densities of the high spin states 
and {\it SF order parameters 
are finite just on a part of the phase boundary curve} 
as long as the phase boundary curve does {\it not} agree with 
that obtained using the PMFA (hereafter,  
we call this part of the phase boundary curve 
as the {\it non-perturbative part}). 
Figure \ref{first} shows the total energy expectation value par site
$\langle H\rangle$ as a function of 
$|g(3)|$ around a MI phase with $N=2$, 
where we see the {\it first-order transition} clearly. 
The high spin states of spin-1 bosons have an essential role in the first-order 
transition: for small $|g(3)|$, the PMFA calculation holds 
and the total energy increases with $|g(3)|$; 
for large $|g(3)|$, $|f(2,2)|$ and $|f(3,3)|$ 
become large and strongly enhance the absolute value of 
the kinetic energy and the total energy decreases with $|g(3)|$; 
for much larger $|g(3)$, the interaction energy within 
$|N=3\rangle$ becomes larger and the total energy again 
increases with $|g(3)|$. 
In summation, the transition between the MI with only the lowest spin state
(which has the lowest antiferromagnetic interaction energy)   
and the SF with higher spin states (which has a large absolute value of 
kinetic energy) can be a first-order one. 

Figure \ref{boundary} shows the chemical potential dependence of 
variational parameters including SF order parameters
($|g(1)|^2$ and $|g(3)|^2$) 
just on the phase boundary around the MI phase with $N=2$. 
These parameters are found to be finite on the non-perturbative part 
and continuously disappear 
at the edges of the non-perturbative part ($\mu/U_0\sim 0.97$ and $1.79$)
where the phase boundary curve agrees with that obtained using the PMFA 
and the transition becomes a second-order one as in the spinless case. 

\begin{figure}
\centerline{\includegraphics[height=2.4in]{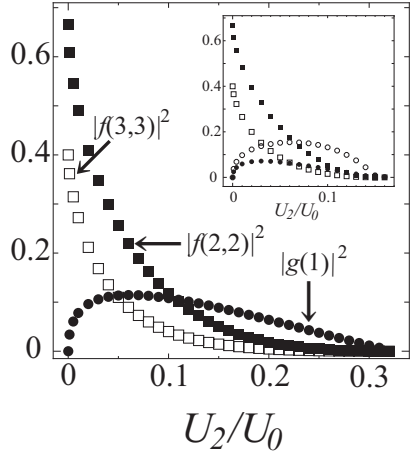}}
%
\caption{
Variational parameters 
just on the Mott lobe of the phase boundary between 
the SF and MI phases with $N=2$
as a function of $U_2$. 
The black circles indicate $|g(1)|^2$,  
the black and white squares indicate $|f(2,2)|^2$ and $|f(3,3)|^2$, 
respectively. 
Here, $|g(3)|=|g(1)|$ within numerical errors. 
The inset indicates the same variational parameters including 
$|g(3)|^2$ (white circles) for $\mu=1.5/U_0$
just on the phase boundary.  
}
\label{crossover}
\end{figure}

The phase boundary curve obtained using the GW around the 
MI phase with $N$ bosons becomes close to 
that obtained using the PMFA for a stronger $U_2$ 
and coincide with it for a finite $U_2$ 
(e.g., $U_2/U_0\sim 0.32$ for $N=2$ and 
$U_2/U_0\sim 0.014$ for $N=3$), 
where the transition becomes a second-order one along the whole 
phase boundary. 
On the other hand, for $U_2=0$, 
the transition also becomes a second-order one because 
the GW has the same form as that of the spinless bosons 
[See above (the sixth paragraph)]. 
Figure \ref{crossover} shows 
the variational parameters just on the Mott lobe of 
the phase boundary between the SF phase and  the 
MI phase with $N=2$ as a function of $U_2/U_0$. 
Here, $\mu$ and $zt$ are determined as a function of $U_2$ 
to obtain the variational parameters just at the Mott lobe. 
We note that the Mott lobe stays on the non-perturbative part 
until the phase diagram perfectly coincides 
with that obtained using the PMFA for $U_2/U_0\sim 0.32$. 
Furthermore, $|g(3)|=|g(1)|$ holds within numerical errors.
While SF order parameters $g(1)$ and $g(3)$ 
disappear for $U_2=0$ and $U_2/U_0\sim 0.32$, 
$|f(2,2)|^2$ and $|f(3,3)|^2$ become larger 
for small $U_2/U_0$ and attain the maximum values for $U_2/U_0=0$. 
This is because  for $U_2/U_0=0$, 
$|N=2\rangle=a_0^{\dagger 2} |0\rangle/\sqrt{2!} 
=(|2,0\rangle+\sqrt{2}|2,2\rangle)/\sqrt{3}$ 
and $|N=3\rangle=a_0^{\dagger 3} |0\rangle/\sqrt{3!} 
=(\sqrt{3}|3,1\rangle+\sqrt{2}|3,3\rangle)/\sqrt{5}$, 
resulting in $|f(2,2)|^2=2/3$ and $|f(3,3)|^2=2/5$. 
The inset of Fig.\ref{crossover} shows 
the $U_2$ dependence of the variational parameters for $\mu/U_0=1.5$ 
just on the phase boundary between the SF 
and MI phases with $N=2$, where $zt$ is determined 
to obtain the variational parameters just at the phase boundary 
as a function of $U_2$. 
The SF order parameters (where $|g(3)|$ is different from $g(1)$) 
continuously disappear for $U_2/U_0\sim0.15$,  
where $\mu/U_0=1.5$ on the phase boundary 
appears away from the non-perturbative part. 

\begin{figure}
\includegraphics[height=2.2in]{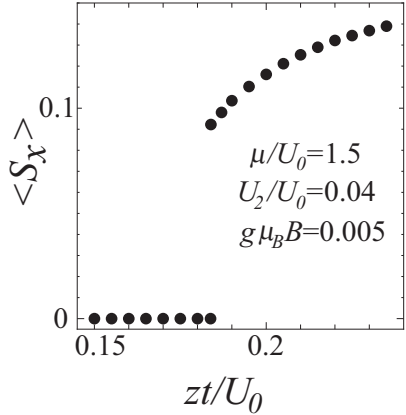}
\caption{$zt/U_0$ dependence of 
$\langle S_x\rangle$ for $U_2/U_0=0.04$ 
and $\mu/U_0=1.5$ under a magnetic field $g\mu_B B=0.005$.
}
\label{mag}
\end{figure}

The first-order transition may be observed 
in future experiments. 
If the lifetimes of locally stable states  
are sufficiently long, one can observe the first-order transition 
through a hysteresis curve because $t/U_0$ and $t/U_2$
can be easily controlled by the laser beam. 
On the other hand, 
the first-order transition may also be observed through 
the response of a spin to a weak magnetic field. 
The magnetization (spin expectation value) under a magnetic field 
may be observed by an experiment such as 
a Stern-Gerlach type time-of-flight measurement 
as discussed in Ref. \cite{plateau}. 
We consider a uniform magnetic field parallel to the $x$-axis \cite{note4}.  
We add the Zeeman coupling $-g\mu_B B\sum_i S_{xi}$ to the Hamiltonian, 
where $g$ is the Lande's $g$-factor of bosons, $\mu_B$ is a Bohr magneton, 
$B$ is the magnetic field, and
$S_{xi}$ is the $x$-component of the spin at site $i$.
We neglect the quadratic Zeeman term because 
a weak magnetic field of the order of mGauss or 
less than mGauss is sufficient \cite{plateau}.
In the GW, the magnetization is also site-independent such 
that $\langle S_{xi}\rangle=\langle S_x\rangle$. 
In a magnetic field, the $S_z=0$ states are not sufficient
to obtain the ground state, and hence, we employ 
the complete set with $S_z=-S,-S+1,\cdot\cdot\cdot,S$ 
in the GW. Figure \ref{mag} shows the $zt/U_0$ dependence of 
$\langle S_x\rangle$ for $U_2/U_0=0.04$ 
and $\mu/U_0=1.5$ under a magnetic field $g\mu_B B=0.005$ \cite{note6}.
We can clearly see that $\langle S_x\rangle$ 
jumps from zero to a finite value for 
$zt/U_0\sim 1.85$, which corresponds to the SF-MI phase boundary
under the magnetic field, and is close to that 
at zero magnetic field $zt/U_0\sim 1.84$. 
In the MI phase, the singlet state at a site 
is stable under a weak magnetic field,  
while in the SF phase, it has a finite spin at a site 
resulting in a finite $\langle S_x\rangle$ under a magnetic field.\cite{note7}  
However, if the transition is continuous, 
$\langle S_x\rangle$ must be a continuous function and should not jump 
at the phase boundary. Hence, this jump of $\langle S_x\rangle$ 
is a unique result of the first-order transition.

We finally note that recent studies \cite{nematic1} have predicted 
possible nematic phases in the MI phase, 
while our approximation results in the lowest spin state in 
the MI phase regardless of the strength of $U_2(>0)$; our study using
the GW cannot include the effects of virtual hopping processes,  
which result in Heisenberg type spin-spin couplings between adjacent sites. 
However, the singlet-nematic phase boundary will be out of the MI phase 
in small densities of ${}^{23}{\mathrm Na}$ atoms 
such as two atoms per site (the case of which is 
 well studied in the present paper) \cite{note5}.
As a matter of course, the relation and/or competition between 
the SF-MI transition and the singlet-nematic transition 
will be an interesting and open subject. 

We acknowledge M. Yamashita, M.W. Jack, 
T. Morishita, S. Watanabe, and K. Kuroki for helpful discussions. 
T. K.  gratefully acknowledges financial support through 
a Grant-in-Aid for the 21st COE Program 
(Physics of Systems with Self-Organization Composed of Multi-Elements).  
S. T. is supported by the Japan Society for the Promotion of Science. 
A part of the numerical calculations was performed at the Supercomputer Center,
Institute for Solid State Physics, University of Tokyo.


\begin{thebibliography}{18}
\bibitem{note1} For simplicity, 
we neglect the weak trapping potential and 
assume that the system is perfectly periodic.
\bibitem{note2} We have numerically checked that the complete set including 
states with finite $S_z$ has the same lowest energy 
as that of the states with only $S_z=0$ at least within the 
states where the number of particles range from $N=0$ to $N=4$.
\bibitem{note0} A GW has 
a unique spin state in an MI because our approximation 
does not include inter-site effective spin coupling 
through virtual hopping processes \cite{nematic1} in the MI.
\bibitem{note5} In refs.\cite{nematic1} it is suggested that 
the singlet-nematic transition occurs for $zt^2/(U_0U_2)=0.5$ in the 
Mott state with two bosons, which corresponds to $zt=0.4$ for $U_2/U_0=0.04$ 
in the case of three dimensional cubic lattices with $z=8$ 
(also see Fig. \ref{phase}).
\bibitem{note6} To obtain this figure, we have employed 
a complete set where the number of bosons range from $N=0$ to $N=4$. 
The states with $N=4$ have very small amplitudes 
($|g(4)|^2\lesssim0.01$ and $|g(4)|^2/|g(3)|^2<0.05$),  
and hence, the numerical convergence is fairly sufficient.
\bibitem{note7} Even if the SF phase has no total spin as a whole system  
(in spite of a finite spin at a site),  
the magnetization will be finite under the magnetic field 
as long as the SF phase has no spin excitation gap.
\end{thebibliography}
\end{document}